\documentclass{99-Styles/MICE} 
\usepackage{textcomp}
\usepackage{lineno}
\usepackage{bibentry}
\nobibliography*
\usepackage{times}
\usepackage{bm}         
\usepackage{amsmath,amssymb,bm,slashed} 
\usepackage{amssymb}    
\usepackage{graphicx}   
\usepackage{verbatim}   
\usepackage{color}      
\usepackage{subfigure}  
\usepackage{hyperref}   
\usepackage{enumerate}

\usepackage{fmtcount}   

\usepackage[square,comma,numbers,sort&compress]{natbib}

\begin{document}

\thispagestyle{empty}

\begin{tabular}{p{0.175\textwidth} p{0.5\textwidth} p{0.225\textwidth}}
  \hspace{-0.8cm}\leftline{\today}                                 &
  \centering{Final (revision 4)}                                   &
  \rightline{ICFA Neutrino Panel 2014(02)} 
\end{tabular}
\vspace{-1.0cm}\\
\rule{\textwidth}{0.43pt}

\begin{center}
  {\bf
    {\LARGE On the complementarity of Hyper-K and LBNF} \\
  }
  \vspace{-0.0cm}
\end{center}

\makeatletter

\newcommand{\bra}[1]{\ensuremath{\langle #1 |}}   
\newcommand{\ket}[1]{\ensuremath{| #1 \rangle}}   
\newcommand{\bigbra}[1]{\ensuremath{\big\langle #1 \big|}}   
\newcommand{\bigket}[1]{\ensuremath{\big| #1 \big\rangle}}   
\newcommand{\amp}[3]{\ensuremath{\left\langle #1 \,\left|\, #2%
                     \,\right|\, #3 \right\rangle}}  
\newcommand{\sprod}[2]{\ensuremath{\left\langle #1 |%
                     #2 \right\rangle}}  
\newcommand{\ev}[1]{\ensuremath{\left\langle #1 %
                     \right\rangle}} 
\newcommand{\ds}[1]{\ensuremath{\! \frac{d^3#1}{(2\pi)^3 %
                     \sqrt{2 E_\vec{#1}}} \,}} 
\newcommand{\dst}[1]{\ensuremath{\! %
                     \frac{d^4#1}{(2\pi)^4} \,}} 
\newcommand{\tr}{\text{tr}}
\newcommand{\sgn}{\text{sgn}}
\newcommand{\diag}{\text{diag}}
\newcommand{\BR}{\text{BR}}

\renewcommand{\vec}[1]{{\mathbf{#1}}}
\renewcommand{\Re}{{\text{Re}}}
\renewcommand{\Im}{{\text{Im}}}
\newcommand{\iso}[2]{{\ensuremath{{}^{#2}}\ensuremath{\rm #1}}}
\newcommand{\eps}{{\ensuremath{\epsilon}}}
\newcommand{\draftnote}[1]{{\bf\color{red} \MakeUppercase{#1}}}
\newcommand{\panm}[1]{{\color{blue} #1}}
\providecommand{\abs}[1]{\lvert#1\rvert}
\providecommand{\norm}[1]{\lVert#1\rVert}

\def\parenbar{\mathpalette\p@renb@r}
\def\p@renb@r#1#2{\vbox{%
  \ifx#1\scriptscriptstyle \dimen@.7em\dimen@ii.2em\else
  \ifx#1\scriptstyle \dimen@.8em\dimen@ii.25em\else
  \dimen@1em\dimen@ii.4em\fi\fi \offinterlineskip
  \ialign{\hfill##\hfill\cr
    \vbox{\hrule width\dimen@ii}\cr
    \noalign{\vskip-.3ex}%
    \hbox to\dimen@{$\mathchar300\hfil\mathchar301$}\cr
    \noalign{\vskip-.3ex}%
    $#1#2$\cr}}}

%
\providecommand{\anmne}{\mbox{$\bar\nu_{\mu} \rightarrow \bar\nu_e$}} 
\providecommand{\nmne}{\mbox{$\nu_{\mu}\rightarrow\nu_e$}} 
\providecommand{\anm}{\mbox{$\bar\nu_\mu$}} 
\providecommand{\nm}{\mbox{$\nu_\mu$}}
\providecommand{\nue}{\mbox{$\nu_e$}} 
\providecommand{\ane}{\mbox{$\bar\nu_e$}} 
\providecommand{\enu}{\mbox{$E_\nu$}}
\providecommand{\piz}{\mbox{$\pi^0 $}}
\providecommand{\pip}{\mbox{$\pi^+$}} 
\providecommand{\pim}{\mbox{$\pi^-$}} 

\parindent 10pt
\pagestyle{plain}
\pagenumbering{arabic}                   
\setcounter{page}{1}

\noindent
The next generation of long-baseline experiments is being designed to
measure neutrino oscillations with a precision substantially better
than that of the present generation in order to:
\begin{itemize}
  \item Search for CP-invariance violation (CPiV) in the lepton
    sector;
  \item Determine the neutrino mass hierarchy; 
  \item Increase substantially the precision with which the
    neutrino-mixing parameters are known; and
  \item Test the three-neutrino-mixing hypothesis (the Standard
    Neutrino Model, S$\nu$M).
\end{itemize}
Two qualitatively different proposals are being considered for
approval:
\begin{itemize}
  \item Hyper-K \cite{Abe:2011ts}, a 560\,kTonne fiducial mass water
    Cherenkov detector located 2.5$^\circ$ off-axis at a distance of
    295\,km from the narrow-band beam produced by an upgraded,
    $\sim 1$\,MW, proton beam at J-PARC; and 
  \item The Long Baseline Neutrino Facility (LBNF)
    \cite{LBNF:iIEB:2014,LBNF:2014}, a 40\,kTonne fiducial mass
    liquid-argon time projection chamber (LAr) illuminated by a new
    wide-band beam produced by a 1.2\,MW proton beam to be built at
    Fermilab.
    The specification of the facility, including the baseline of
    $1\,300$\,km, and the choice of detector technology will take
    advantage of the design studies already performed for
    LBNE \cite{Adams:2013qkq} and
    LBNO \cite{LAGUNA-LBNO-Docs:2015,Agarwalla:2014ura,Agarwalla:2014tca,::2013kaa}.
\end{itemize}
Each project is ambitious and requires an investment significantly
larger than any previous single investment in a neutrino experiment.
This document outlines the complimentarity between Hyper-K and LBNF.

\section{The experiments}

The critical features of the Hyper-K proposal that determine the
physics performance include:
\begin{itemize}
  \item{\it Accelerator-based oscillation measurements:}
    The relatively short baseline implies small matter effects.
    This reduces the effect of correlations among the oscillation
    parameters in, for example, searches for CPiV.
    The off-axis, narrow-band beam peaks at $\sim 600$\,MeV.
    The suppressed high-energy tail leads to a reduced neutral-current
    background in $\parenbar{\nu}_e$-appearance samples.
  \item{\it Non-accelerator-based neutrino measurements:}
    The large atmospheric-neutrino data set will allow the mass
    hierarchy to be determined and may be sensitive to new phenomena.
    In the event of a nearby supernova, a large sample of
    $\bar{\nu}_e$ events would be recorded.
  \item{\it Proton decay:}
    Hyper-K is unique in its ability to extend the limits on the
    majority of proton decay modes by an order of magnitude.
  \item{\it Accelerator and detector R\&D:}
    Incremental developments to proton source, target and horn are
    required for the beam power of $\sim 1$\,MW to be delivered.
    An R\&D programme is underway to reduce the cost of photosensors
    with the required collection efficiency.
    The T2K near detector programme will provide valuable constraints
    on neutrino flux and neutrino-interaction rates.
    The development of a dedicated near detector as part of the
    Hyper-K programme is essential for the experiment to fulfil its
    potential.
\end{itemize}

\noindent
The critical features of the LBNF proposal that determine the physics
performance include:
\begin{itemize}
  \item{\it Accelerator-based oscillation measurements:}
    The long baseline yields a significant matter effect that can be
    used to determine the mass hierarchy.
    The excellent energy resolution and background rejection offered
    by the LAr technique allows the first and second oscillation
    maxima to be studied.
    The energy of the LBNF neutrino beam is sufficient for the
    reaction $\parenbar{\nu}_\tau N \rightarrow \tau^\pm X$ to occur
    at an appreciable rate and the LAr technique has the potential to
    isolate samples of $\parenbar{\nu}_\tau$ of significant size.
    This will provide an important opportunity to test the S$\nu$M.
  \item{\it Non-accelerator-based neutrino measurements:}
    LBNF would accumulate a large ``high-resolution'' atmospheric
    neutrino sample.
    The LAr detector is sensitive to the $\nu_e$ (rather than the
    $\bar{\nu}_e$) component of the supernova flux, which contains
    information regarding the ``neutronization burst'' ($p+e^-\to
    n+\nu_e$) that is expected to take place in the early stages of
    the explosion.
  \item{\it Proton decay:}
    The LAr detector, which will be substantially larger than any of
    those built to date, will offer the opportunity to search for
    proton-decay modes that are ``preferred'' by supersymmetric 
    models.
  \item{\it Accelerator and detector R\&D:}
    The total fiducial mass of the LAr detector will be implemented
    using a modular approach.
    An R\&D programme to develop the necessary techniques is
    underway.
    The existing LAr-based program that includes ArgoNeuT, MicroBooNE,
    the LBNE 35\,Ton prototype, CAPTAIN and LArIAT and the R\&D
    projects developed at the CERN Neutrino Platform, will provide
    important information on neutrino-argon interactions, detector
    response, and reconstruction algorithms.
    A highly segmented near detector is essential for the
    facility to fulfil its potential and is under development.
    The proton-beam power of 1.2\,MW will be delivered by the Proton
    Improvement Plan II upgrade to the Fermilab accelerator complex.
    This requires a 800\,MeV, superconducting H$^-$ linac and a new
    high-power pion-production target.
\end{itemize}

\section{The qualitative case for both experiments}

The accelerator based programmes of both LBNF and Hyper-K have been
optimised at the same $L/E$ but the baselines, $L$, and energies, $E$,
differ by almost a factor of 5.
In each experiment, the large atmospheric-neutrino sample will extend
substantially the range of $L$ and $E$ that can be studied.
The two experiments have similar sensitivities to CPiV and will be
able to measure the mixing angles with comparable precision.
The longer baseline allows LBNF to determine the mass hierarchy.
If the two experiments progress on technically-limited schedules they
will compete to discover CPiV.
Given the challenging nature of the measurement and the importance of
the discovery, independent confirmation by a qualitatively different
experiment is likely to be essential.
 
The differing degree to which the matter effect modifies the
oscillation probabilities at Hyper-K and LBNF may be exploited to
break parameter degeneracies.
The different detector technologies and beam energies imply that
neutrino scattering in the two experiments is dominated by different
interaction processes and hence different event topologies.
In addition, the details of the hadronic processes by which the two
nenutrino beams are generated are different.
Therefore, the systematics are quite different at Hyper-K and LBNF.

For a given parameter set, the S$\nu$M specifies that the oscillation
probabilities are functions of $L/E$.
Searches for non-standard phenomena can therefore be made by
exploiting the fact that Hyper-K and LBNF have the same $L/E$ but
different $L$ and $E$. 
To put the S$\nu$M to the test requires precise measurements of
several observables, including $\parenbar{\nu}_e$ (and, if possible,
$\parenbar{\nu}_\tau$) appearance from a $\parenbar{\nu}_{\mu}$ beam
at different values of $L$ and $E$, comparing neutrino data with
antineutrino data and comparing accelerator-based measurements with
measurements of $\bar{\nu}_e$ disappearance at reactors.

Hyper-K offers the opportunity to extend the sensitivity to
proton decay significantly in several modes, while the LAr detector at
LBNF will probe new decay channels that are not accessible to water
Cherenkov detectors and will provide nearly background-free searches
for other important channels. 
To understand the mechanisms of supernova explosion requires accurate
measurements of the $\nu_e$ and $\bar{\nu}_e$ fluxes, along with
some neutral current data (which is sensitive to the flux of
$\parenbar{\nu}_{\mu,\tau}$).
These measurements can not be made with Hyper-K or LBNF alone.

\section{The need to quantify added value}

Detailed simulations are required in order to quantify the
complimentarity between Hyper-K and LBNF.
Some representative questions that need quantitative studies are:
\begin{itemize}
  \item{\it Searching for CP-invariance violation:}
    the degree to which the combined data set enhances the sensitivity 
    of searches for CP-invariance violation;
  \item{\it Lifting degeneracies:}
    the degree to which the combined data set reduces the number
    of viable regions of multi-parameter space;
  \item{\it Improved precision:}
    the degree to which fits to the combined data set, which assumes
    the validity of the S$\nu$M,  will improve the precision of the
    parameter determination; and
  \item{\it Testing the S$\nu$M framework:}
    the degree to which the combined data set enhances the coverage of
    the non-standard-neutrino-model parameter space.
\end{itemize}
The benefit that will accrue from the parallel implementation of these
complementary experiments should be quantified at an early stage.

\bibliographystyle{99-Styles/utphys}
\bibliography{Concatenated-bibliography}

\clearpage
\appendix
\section{The ICFA Neutrino Panel}
\label{App:ICFA-nu-Panel}

ICFA established the Neutrino Panel with the mandate
\cite{ICFAnuPanel:Mandate:2013}: 
\begin{quote}
  {\it To promote international cooperation in the development of the
    accelerator-based neutrino-oscillation program and to promote
    international collaboration in the development a neutrino factory
    as a future intense source of neutrinos for particle physics
    experiments.
  }
\end{quote}
The membership of the Panel agreed by ICFA at its meeting in February
2013 is shown in table \ref{Tab:PanelMembers}.
The terms of reference for the panel \cite{ICFAnuPanel:ToR:2013} may
be found on the Panel's WWW site \cite{ICFA:nuPanelWWWSite}.
\begin{table}[h]
  \caption{Membership of the ICFA Neutrino Panel.}
  \label{Tab:PanelMembers}
  \begin{center}
    \begin{tabular}{|l|l|}
      \hline
      {\bf Name}      & {\bf Institution}                      \\
      \hline
      J. Cao          & IHEP/Beijing                           \\
      A. de Gouv\^ea  & Northwestern University                \\
      D. Duchesneau   & CNRS/IN2P3                             \\
      R. Funchal      & University of Sao Paulo                \\
      S. Geer         & Fermi National Laboratory              \\
      S.B. Kim        & Seoul National University              \\
      T. Kobayashi    & KEK                                    \\
      K. Long (chair) & Imperial College London and STFC       \\
      M. Maltoni      & Universidad Automata Madrid            \\
      M. Mezzetto     & University of Padova                   \\
      N. Mondal       & Tata Institute for Fundamental Resarch \\
      M. Shiozawa     & Tokyo University                       \\
      J. Sobczyk      & Wroclaw University                     \\
      H. A. Tanaka    & University of British Columbia and IPP \\
      M. Wascko       & Imperial College London                \\
      G. Zeller       & Fermi National Accelerator Laboratory  \\
      \hline
    \end{tabular}
  \end{center}
\end{table}

\end{document}